\documentclass{article}
\usepackage{graphicx} 
\usepackage{amsmath}
\usepackage{amssymb}
\usepackage{amsthm}
\usepackage{color}
\allowdisplaybreaks
\usepackage[backend=biber, sorting=none]{biblatex}
\usepackage{csquotes}
\usepackage{authblk}
\usepackage{authblk}

\theoremstyle{definition}

\theoremstyle{remark}

\title{On induced L-infinity action of diffeomorphisms on Cochains}
\author[1]{Andrey Losev}
\author[2]{Dmitrii Sheptunov}
\author[3]{Xin Geng}
\affil[1]{Shanghai Institute for Mathematics and Interdisciplinary Sciences}
\affil[2]{Moscow Institute of Physics and Technology}
\affil[3]{University of Science and Technology of China}
\date{\today}

\begin{document}

\maketitle	
\begin{abstract}
  One of the approaches to quantum gravity is to formulate it in terms of De Rham algebra, choose a triangulation of space-time, and replace differential forms by cochains (that form a finite dimensional vector space). The key issue of general relativity is the action of diffeomorphisms of space-time on fields.
  In this paper, we induce the action of diffeomorphisms on cochains by homotopy transfer (or, equivalently, BV integral) that leads to a $L_{\infty}$ action.
  We explicitly compute this action for the space-time being an interval, a circle, and a square.
\end{abstract}

\section{Introduction}

The quantum theory of gravity is one of the main goals of theoretical physics. The main problem is in ultraviolet divergences that are present due to the infinite dimension of the space of differential forms
on the smooth manifold. Therefore, to quantize gravity in the spirit of Feynman integral, we should modify the space of differential forms. In (\cite{hu2016feynman})
we proposed the replacement of the differential forms by A-infinity algebra, that was called Feynman geometry.

The particular class of Feynman geometries may be obtained in the following construction.
Consider triangulation of a manifold. Decompose the space of differential forms as complexes as a direct sum of an infinite-dimensional complex of differential forms that have zero integrals against all chains
of the triangulation, and a finite dimensional complex of cochains dual to such chains of the triangulation. The latter would be a replacement of the algebra of differential forms.
In works of Mnev (\cite{mnev2006simplicial, mnev2008discrete}) and Sullivan (\cite{lawrence2006free}) it was shown how wedge product on differential forms leads to higher products of cochains in the procedure known as homotopic transfer, which in mathematical physics
is just a BV integral against a Lagrangian submanifold determined by the homotopy.

Since we are planning to study gravity the important issue is the general covariance. In particular, we should know how diffeomorphisms are acting on cochains.
The hint is that we know that diffeomorphisms are acting on differential forms by Lie derivative. 
Thus, in this paper, we will apply the BV integral (homotopic transfer) to get the action of vector fields on manifold on cochains. It is not surprising that we will get not the representation
of the Lie algebra but L-infinity representation. This construction in the context of transferring the action of supersymmetry to cohomology was earlier done in (\cite{alexandrov2007pure}).

We believe that explicit formulas for such representation would be an important stepping stone in understanding of quantum gravity in this approach.

The paper is organized as follows. In section 2 we review relation between $L_\infty$ representations and solutions to classical master equations (CME) of the special form. We also recall in this section that BV integral leads to induction of  $L_\infty$ representation (also known as homotopic transfer). These results are not new, but we present them for completeness.

In section 3 we proceed to explicit calculation of effective BV-action. To begin with, we consider an interval as a manifold. Further, in sections 4 and 5, explicit formulas for the action in the case of a circle and a square are obtained.

\section{BV formalism, $L_\infty$ module structure, BV integral and homotopic transfer}

\subsection{How to understand $L_{\infty}$ module}
The concept of $L_\infty$ module (\cite{kontsevich2003deformation}) could be easily understood by analogy with the concept of a module over a Lie algebra. The latter may be explained as follows. Take a Lie algebra $L$, a vector space $M$
(considered as an abelian Lie algebra) and study
$L\oplus M$. Now deform the Lie algebra structure on $L \oplus M$ considered as a vector space keeping the structure constants of $L$ .
This would lead to structure constants 
$$
T : \; \;L \otimes M \rightarrow M
$$
and condition that $T$ gives a structure of the Lie algebra module is equivalent to Jacobi identity for the deformed Lie algebra.

If we apply the same procedure to $L_{\infty}$ algebra, namely, extend it by a complex $M$ and deform, we will get a set of maps
$$
T_p : \; \; \bigwedge \, ^p L \rightarrow End(M)
$$
subject to quadratic relations that follow from 
the $L_{\infty}$ relations.

\subsection{BV formalism}
A convenient way to write down these relations and to operate with them is provided by BV formalism (see \cite{schwarz1993semiclassical}).

In this formalism we study a supermanifold equipped with the odd simplectic structure and a function 
$S_{BV}$ on it, that is called BV action. 
$S_{BV}$ should satisfy quantum master equation, but
in this work we will restrict ourselves by its classical limit, that we will write explicitly below.

For our purpose it is enough to consider
as a supermanifold $T^*[1]W$, where $W$ is a supervector space and simplectic structure is a canonical one.
In BV language linear coordinates on $W$ are called fields ( and we will denote them as $\eta^a$) and dual linear coordinates on the fiber are called antifields and we will denote them as $\eta_a^*$.

The classical master equation (CME) takes the form

\begin{equation}
 \sum_{a=1}^{dim V} (-1)^{|\eta^a|} \Bigg( \frac{\partial S_{BV}}{\partial \eta^a} \frac{\partial S_{BV}}{\partial \eta_a^*} + \frac{\partial S_{BV}}{\partial \eta_a^*} \frac{\partial S_{BV}}{\partial \eta^a} \Bigg) = 0
\end{equation}
where $|\eta^a|$ is parity of $\eta^a$.

Using BV formalism is useful because one can reformulate the concept of $L_{\infty}$ algebra and
the concept of $L_{\infty}$ module as a particular solutions to CME.
Moreover, the BV formalism contains the procedure of
BV integral allowing to get new solutions of CME from the old ones.

Namely,
if $W=W_1\oplus W_2$, and $\mathcal{L}$ is the Lagrangian submanifold in $T^*[1]W_1$, then the integral
\begin{equation}
    \int_{\mathcal{L}} \exp( S_{BV}/\hbar)=
    \exp( S_{BV}^{ind}(\eta_2,\eta_2^*,\hbar)/\hbar)
\end{equation}
produces the induced BV action $S_{BV}^{ind}(\hbar)$
that satisfies quantum master equation.
For our purpose it is enough to consider the value of
$S^{ind}$ at $\hbar=0$ that gives a solution to CME on the space 
$T^*[1]W_2$.
Note, that 
\begin{equation}
    S_{BV}^{ind}(\eta_2, \eta_2^*,0)=extr_{\mathcal{L}}
    S_{BV} (\eta, \eta^*)
\end{equation}

\subsection{$L_{\infty}$ algebra and $L_{\infty}$ module from the BV point of view}

Consider the Lie algebra $L$ with a basis \{ $t_a$ \}
and structure constants $f_{ab}^{e}$:
\begin{equation}
[t_a, t_b]=f_{ab}^{d} t_d
\end{equation}
Consider $L[1]$, it means that $L[1]$ is an odd space with odd coordinates $c^a$.

Correspondingly, dual coordinates on the fiber $T^*[1] (L[1])$ are even and according to our conventions we denote them as $c_a^*$

It is easy to show that CME for the action
\begin{equation}
 S_L=f_{ab}^d c^a c^b c_d^*   
\end{equation}
is equivalent to Jacobi identity for the structure constants $f_{ab}^{d}$.

Consider an (even) vector space $M$ with the basis $e_i$ and linear coordinates $\varphi^i$. The fiber $T^*[1]M$ has odd dual coordinates $\varphi_i^*$. It is equally easy to check that $M$ is a module over Lie algebra iff the
map 
\begin{equation}
 L \rightarrow End(M),  t_a \mapsto T_{aj}^{i}  
\end{equation}
is such that
\begin{equation}
 S_M= c^a T_{aj}^{i} \varphi_i^* \varphi^j + \frac{1}{2} f_{ab}^d c^a c^b c_d^*
\end{equation}
solves CME.

Note, that the latter case may be considered as a particular case of the former one for the superalgebra obtained by extension of $L$ by the odd module $M$.

It is instructive to notice that the $L_{\infty}$ algebra structure, i.e. the collection of maps
\begin{equation}
 \bigwedge L \rightarrow L, \; \; \; 
  t_{a_1}\wedge \ldots \wedge t_{a_p} \mapsto 
  f_{a_1 \ldots a_p}^b t_b
\end{equation}
satisfying quadratic relations can be packed into a solution to CME for
\begin{equation}
 S_{L_\infty}= \sum\limits_{p=1}^n f_{a_1 \ldots a_p}^b c^{a_1} \ldots c^{a_p} c_b^*   
\end{equation}

 Now we are ready to explain the concept of $L_{\infty}$ module in BV formalism.

 Let us replace a supervector space $M$ by a complex with
 the differential $Q$ and introduce the matrix $Q_{i}^{j}$ as follows:
 \begin{equation}
     Q(e_i)=Q_{i}^{j}e_j, \; \; \; Q^2=0
 \end{equation}

Define a map 
\begin{equation}
    \bigwedge L \rightarrow End(M), \; \; 
    t_{a_1}\wedge \ldots \wedge t_{a_p} \mapsto 
    (T_{a_{1}\ldots a_{p}})_{i}^{j}
\end{equation}

Then one can show that this map gives a stucture of $L_{\infty}$
module iff the follwing action $S_{M_{\infty}}$ solves CME:

\begin{equation}
S_{M_{\infty}}=\varphi^*_i Q^i_j \varphi^j + \sum_{p=1}^n\frac{1}{p!}c^{a_1}\cdots c^{a_p}(T_{a_1\cdots a_p})^i_j \varphi^*_i \varphi^j+\frac{1}{2}f_{ab}^d c^ac^b c^*_d \label{BV_action}
\end{equation}

Like in the case of ordinary module over a Lie algebra, the structure of $L_{\infty}$ module may be considered as a special case of $L_{\infty}$ algebra
on the space $L \oplus M$.

\subsection{Induction of the $L_{\infty}$ module by contraction of the acyclic subcomplex in the BV formalism}

Kadeishvili-like theorems (\cite{kadeishvili1980theory}) state that by contraction of the acyclic subcomplex we can pass from the Lie algebra to $L_{\infty}$ algebra. Here we will describe how in works in the particular case of $L$ modules and $L_{\infty}$ modules. 

Let us apply the BV induction procedure for the Lie algebra acting on a complex $M$ with the action commuting with the differential $Q$ in the complex:
\begin{equation}
    [T_a, Q]=0
\end{equation}

Let $M$ be a direct sum of two subcomplexes:
\begin{equation}
    M=M_Z\oplus M_A,
\end{equation}
such that $M_A$ is acyclic, and $h$ is the contracting homotopy:
\begin{equation}
hQ+Qh=Proj_{M_A},  \; \; h^2=0 , \; \; h M_Z=0   \label{homotopy_cond} 
\end{equation}

Take
\begin{equation}
 W_1=M_A,  \; \;  W_2=L[1] \oplus M_Z    
\end{equation}
and consider linear Lagrangian submanifold given by equations
\begin{equation}
    h w=0 ; \; \; h^* w^*=0
\end{equation}
where 
$h^*$ is a conjugated homotopy operator on the dual space. 

Since $S_M$ is quadratic polynomial in $w_A \in M_A$
and $w_A^* \in M_A^*[1]$, it is easy to compute its critical value on the Lagrangian submanifold.

It is given by 
\begin{equation}
    S^{ind}= <w_Z^* , Q w_Z> + \frac{1}{2} f_{ab}^d c^a c^b c_d^*   +
    \sum_{k=1}^{\infty} c^{a_1} \ldots c^{a_k}
   <w_Z^* , T_{a_1} h \ldots h T_{a_k} w_Z> \label{induced_action}
\end{equation}
where $w_Z \in M_Z$ and $w_Z^* \in M_Z^*[1]$. If we introduce an universal action operator
$$
T_c=c^a T_a \label{generators_sum}
$$
the formula above takes the simple form
\begin{equation}
    S^{ind}= <w_Z^* , Q w_Z> + \frac{1}{2} f_{ab}^d c^a c^b c_d^*   +
    \sum_{k=1}^{\infty} 
   <w_Z^* , \underbrace{T_c h \ldots h T_c}_\text{k times} w_Z> 
\end{equation}

Thus we have an induced structure of $L_\infty$ module also known as homotopy transfer, see for example (\cite{arvanitakis2022homotopy}).

\subsection{Application to the action of diffeomorphisms on cochains}

In our approach to discretized gravity, we take as a module $M$ the space $\Omega_X$ of differential forms on a manifold $X$, as a differential - De Rham operator $d$ on these forms, as a Lie algebra - the algebra of diffeomorphisms on $X$.

Now we proceed to discretization. We consider a triangulation of $X$, and take as $M_A$ the subspace $\Omega_A$ of differential forms $\omega$ having zero integrals against all the chains $z$ of the triangulation:
\begin{equation}
 \int_z \omega=0 
\end{equation}

As a second summand we take differential forms $\Omega_Z$ dual to chains of the triangulation. Such
choice is not unique. Whitney considered polynomial
forms. However, in this paper we allow more general choices,restricted by condition that
$$
d \Omega_Z \in \Omega_Z
$$

Applying the construction described above, we get $L_{\infty}$ module structure on $\Omega_Z$. In the following sections we will study explicit examples.

\section{Interval}

In this section we consider the manifold $X = [0, 1]$ with the coordinate $t$, $0 \le t \le 1$. The corresponding Lie algebra $L = Vect([0, 1])$ is spanned by
\begin{equation}
    v_k = t^k \frac{d}{dt}, k \in \mathbb{Z}, k \ge 0
\end{equation}
The structure constants can be written in this basis as
\begin{align}
    & [v_k, v_l] = (l - k) t^{k + l - 1} \frac{d}{dt} = (l - k) v_{k+l-1} \\
    & f_{kl}^m = (l - k) \delta_{k+l-1}^m
\end{align}
We will also use the notation like in (\ref{generators_sum})
\begin{align}
    & v_c = \sum\limits_k c^k v_k = \sum\limits_k c^k t^k \frac{d}{dt} = \nu_c(t) \frac{d}{dt}
\end{align}

The action of $v_c \in L[1]$ on $\Omega_X$ is given by the Lie derivative which we denote $\mathcal{L}_{v_c}$. So the action (\ref{BV_action}) takes the form
\begin{equation}
    S = \langle \varphi^*, d\varphi \rangle + (-1)^{|\varphi^*|} \langle \varphi^*, \mathcal{L}_{v_c} \varphi \rangle + \frac{1}{2} f_{ab}^d c^a c^b c_d^*
\end{equation}

We construct the triangulation of $[0, 1]$ as follows. Choose $n + 1$ points $t_0 < t_1 <...< t_n$ such that $t_0 = 0, t_n = 1$. Using this we set the space of chains
\begin{equation}
    Z = Span \langle \alpha^0 = [t_0],..., \alpha^n = [t_n], \beta^0 = [t_0, t_1],..., \beta^{n-1} = [t_{n-1}, t_n] \rangle
\end{equation}


Further consider $n$ functions $\alpha_1,..., \alpha_n$ with the condition $\alpha_j(t_i) = \delta_j^i$ and define $\alpha_0 = 1 - \sum\limits_{k=1}^n \alpha_k$. 1-forms are introduced as
\begin{equation}
    \beta_j = d \Bigg( \sum\limits_{k=j+1}^n \alpha_k \Bigg), 0 \le j \le n-1
\end{equation}
Note that:
\begin{equation}
    \int\limits_{t_i}^{t_{i+1}} \beta_j = \int\limits_{t_i}^{t_{i+1}} d \Bigg( \sum\limits_{k=j+1}^n \alpha_k \Bigg) = \sum\limits_{k=j+1}^n \alpha_k(t_{i+1}) - \alpha_k(t_i) = \sum\limits_{k=j+1}^n \delta_k^{i+1} - \delta_k^i = \delta_j^i
\end{equation}
Hence we get $ \langle \alpha^i, \alpha_j \rangle = \langle \beta^i, \beta_j \rangle = \delta_j^i$, $ \langle \alpha^i, \beta_j \rangle = \langle \beta^i, \alpha_j \rangle = 0$, where $\langle , \rangle$ is the pairing between chains and differential forms given by integration, and the space of forms dual to chains can be written as:
\begin{equation}
    \Omega_Z = Span(\alpha_0,..., \alpha_n, \beta_0,...,\beta_{n-1})
\end{equation}
Therefore, fields and antifields are
\begin{align}
    & \varphi_Z = \sum\limits_{i=0}^n \alpha_i \varphi^i + \sum\limits_{j=0}^{n-1} \beta_j \psi^j \\
    & \varphi_Z^* = \sum\limits_{i=0}^n \varphi_i^* \alpha^i + \sum\limits_{j=0}^{n-1} \psi_j^* \beta^j
\end{align}

To construct the $L_\infty$ module structure on $\Omega_Z$, we need to provide a homotopy $h$ satisfying (\ref{homotopy_cond}). We write this condition in the form $dh + hd = id - P_Z$, where the projector $P_Z: \Omega_X \to \Omega_Z$ is defined as
\begin{align}
    P_Z(g(t)) = \sum\limits_j g(t_j) \alpha_j 
\end{align}
for a 0-form $g(t)$ and
\begin{align}
    P_Z(f(t) dt) = \sum\limits_j \Bigg( \int\limits_{t_j}^{t_{j+1}} f(\tau) d\tau \Bigg) \beta_j
\end{align}
for a 1-form $f(t)dt$.

Let us set now
\begin{align}
    h(f(t) dt) = \int\limits_0^t f(\tau) d\tau - \sum\limits_j \Bigg( \int\limits_{t_j}^{t_{j+1}} f(\tau) d\tau \Bigg) \int\limits_0^t \beta_j
\end{align}
and check the relation $dh + hd = id - P_Z$ for 0-forms:
\begin{align}
    (dh + hd)g(t) & = h(g'(t)dt) = \int\limits_0^t g'(\tau) d\tau - \sum\limits_j \Bigg( \int\limits_{t_j}^{t_{j+1}} g'(\tau) d\tau \Bigg) \int\limits_0^t \beta_j = \\
    & = g(t) - g(0) - \sum\limits_j \Big( g(t_{j+1}) - g(t_j) \Big) \int\limits_0^t d \Bigg( \sum\limits_{k=j+1}^n \alpha_k \Bigg) = \\
    & = g(t) - g(0) - \sum\limits_j \Big( g(t_{j+1}) - g(t_j) \Big) \sum\limits_{k=j+1}^n \alpha_k(t) = \\
    & = g(t) - g(0) - \sum\limits_{k=1}^n \alpha_k(t) \sum\limits_{j=0}^{k-1} \Big( g(t_{j+1}) - g(t_j) \Big) = \\
    & = g(t) - g(0) - \sum\limits_{k=1}^n \alpha_k(t) \Big( g(t_k) - g(t_0) \Big) = \\
    & = g(t) - g(t_0) (1 - \sum\limits_{k=1}^n \alpha_k) - \sum\limits_{k=1}^n g(t_k) \alpha_k = (id - P_Z) g(t)
\end{align}
and for 1-forms:
\begin{align}
    (dh + hd) f(t)dt & = d \Bigg[ \int\limits_0^t f(\tau) d\tau - \sum\limits_j \Bigg( \int\limits_{t_j}^{t_{j+1}} f(\tau) d\tau \Bigg) \int\limits_0^t \beta_j \Bigg] = \\
    & = f(t)dt - \sum\limits_j \Bigg( \int\limits_{t_j}^{t_{j+1}} f(\tau) d\tau \Bigg) \beta_j = (id - P_Z) f(t)dt
\end{align}

Finally, the induced action (\ref{induced_action}) becomes
\begin{equation}
    S^{ind} = \langle \varphi^*_Z, d\varphi_Z \rangle + (-1)^{|\varphi_Z^*|} \langle \varphi^*_Z, \mathcal{L}_{v_c} \varphi_Z \rangle + \langle \varphi^*_Z, \mathcal{L}_{v_c} h \mathcal{L}_{v_c} \varphi_Z \rangle + \frac{1}{2} f_{ab}^d c^a c^b c_d^*
\end{equation}
All the terms containing more than one operator $h$ vanish since $h$ lowers form's degree by 1.

To compute the induced action we introduce notations
\begin{align}
    & \beta_j^i = \sum\limits_{k=j+1}^n \alpha'_k(t_i) \\
    & \alpha^j \nu_c = \nu_c(t_j) = \nu_c^j \\
    & \alpha^j \nu'_c = \nu'_c(t_j) = \nu'^j_c
\end{align}
What remains is to calculate the distinct components of the expression for the induced action
\begin{align}
    & \langle \beta^j, d\alpha_i \rangle = \int\limits_{t_j}^{t_{j+1}} d\alpha_i = \alpha_i(t_{j+1}) - \alpha_i(t_j) = \delta_i^{j+1} - \delta_i^j \\
    & \langle \alpha^i, \mathcal{L}_{v_c} \alpha_j \rangle = \alpha^i \big( \nu_c(t) \frac{d \alpha_j}{dt} \big) = \nu_c(t_i) \alpha'_j(t_i) = \nu_c^i (\beta_{j-1}^i - \beta_j^i) \\
    \begin{split}
        \langle \beta^i, \mathcal{L}_{v_c} \beta_j \rangle & = \beta^i d \Bigg( \nu_c(t) \sum\limits_{k=j+1}^n \alpha'_k \Bigg) = \int\limits_{t_i}^{t_{i+1}} d \Bigg( \nu_c(t) \sum\limits_{k=j+1}^n \alpha'_k \Bigg) \\
        & = \sum\limits_{k=j+1}^n \nu_c(t_{i+1}) \alpha'_k(t_{i+1}) - \nu_c(t_i) \alpha'_k(t_i) = \nu_c^{i+1} \beta_j^{i+1} - \nu_c^i \beta_j^i \\
    \end{split} \\
    \begin{split}
        \langle \alpha^i, \mathcal{L}_{v_c} h \mathcal{L}_{v_c} \beta_j \rangle & = \alpha^i \mathcal{L}_{v_c} h d \Bigg( \nu_c(t) \sum\limits_{k=j+1}^n \alpha'_k \Bigg) = \alpha^i \mathcal{L}_{v_c} (id - P_Z) \Bigg( \nu_c(t) \sum\limits_{k=j+1}^n \alpha'_k \Bigg) = \\
        & = \alpha^i \mathcal{L}_{v_c} \Bigg( \nu_c(t) \sum\limits_{k=j+1}^n \alpha'_k - \sum\limits_m \nu_c^m \beta_j^m \alpha_m \Bigg) = \\
        & = \nu_c^i \nu'^i_c \beta_j^i - \nu_c^i \sum\limits_m \nu_c^m \beta_j^m (\beta_{m-1}^i - \beta_m^i) \\
    \end{split}
\end{align}

Thus, the resulting $L_\infty$ module structure is
\begin{align}
    \begin{split}
        S^{ind} & = \psi_i^* (\delta_j^{i+1} - \delta_j^i) \varphi^j - \varphi_i^* \nu_c^i (\beta_{j-1}^i - \beta_j^i) \varphi^j + \psi_i^* (\nu_c^{i+1} \beta_j^{i+1} - \nu_c^i \beta_j^i) \psi^j + \\
        & + \varphi_i^* \Big( \nu_c^i \nu'^i_c \beta_j^i - \nu_c^i \sum\limits_m \nu_c^m \beta_j^m (\beta_{m-1}^i - \beta_m^i) \Big) \psi^j + \frac{1}{2} f_{ab}^d c^a c^b c_d^*
    \end{split}
\end{align}

\section{Circle}

In this section we consider the manifold $X = S^1$ with the coordinate $0 \le t < 2\pi$ and the Lie algebra $L = Vect(S^1)$. Let
\begin{equation}
    v_k = e^{i k t} \frac{d}{dt}, k \in \mathbb{Z}
\end{equation}
be a basis in $L = Vect(S^1)$.
Then, the corresponding structure constants are
\begin{align}
    & [v_k, v_l] = i (l - k) e^{i(m+n)t} \frac{d}{dt} = i (l - k) v_{k+l} \\
    & f_{kl}^m = i (l - k) \delta_{k+l}^m
\end{align}

Let us use the following notation for the subsequent computations
\begin{align}
    & v_c = \sum\limits_k c^k v_k = \sum\limits_k c^k e^{ikt} \frac{d}{dt} = \nu_c(t) \frac{d}{dt}
\end{align}

The action (\ref{BV_action}) takes the form analogous to the interval case
\begin{equation}
    S = \langle \varphi^*, d\varphi \rangle + (-1)^{|\varphi^*|} \langle \varphi^*, \mathcal{L}_{v_c} \varphi \rangle + \frac{1}{2} f_{ab}^d c^a c^b c_d^*
\end{equation}


The triangulation of $S^1$ is constructed as follows. Choose $n$ points $t_0=0,..., t_{n-1}$ and 1-forms $\beta_0,..., \beta_{n-1}$ such that $\int\limits_{t_i}^{t_{i+1}} \beta_j = \delta_j^i$. Denote $\beta_j^i := \vartheta_j(t_i)$, where $\beta_j = \vartheta_j(t) dt$. Define the 0-forms:
\begin{align}
    & \alpha_0(t) = 1 + \int\limits_0^t(\beta_{n-1} - \beta_0) \\
    & \alpha_j(t) = \int\limits_0^t(\beta_{j-1} - \beta_j), j \ne 0
\end{align}
One can check that $\alpha_j(t_i) = \delta_j^i$.

Then, the spaces of chains and cochains are
\begin{align}
    & Z = Span \langle \alpha^0 = [t_0],..., \alpha^{n-1} = [t_{n-1}], \beta^0 = [t_0, t_1],..., \beta^{n-1} = [t_{n-1}, t_0] \rangle \\
    & \Omega^\bullet_Z = Span \langle \alpha_0,..., \alpha_{n-1}, \beta_0,..., \beta_{n-1} \rangle
\end{align}

The corresponding fields and antifields are
\begin{align}
    & \varphi_Z^* = \sum\limits_{i=0}^{n-1} \varphi_i^* \alpha^i + \sum\limits_{j=0}^{n-1} \psi_j^* \beta^j \\
    & \varphi_Z = \sum\limits_{i=0}^{n-1} \alpha_i \varphi^i + \sum\limits_{j=0}^{n-1} \beta_j \psi^j
\end{align}

As mentioned earlier in section 3, it is necessary to provide $h$ satisfying the condition $dh + hd = id - P_Z$, where the projector $P_Z: \Omega_X \to \Omega_Z$ is defined in this case as
\begin{align}
    P_Z(g(t)) = \sum\limits_j g(t_j) \alpha_j 
\end{align}
for a 0-form $g(t)$ and
\begin{align}
    P_Z(f(t)dt) = \sum\limits_j \Bigg( \int\limits_{t_j}^{t_{j+1}} f(\tau) d\tau \Bigg) \beta_j
\end{align}
for a 1-form $f(t)dt$.

Let us set the homotopy $h$
\begin{align}
    h(f(t)dt) = \int\limits_0^t f(\tau) d\tau - \sum\limits_j \Bigg( \int\limits_{t_j}^{t_{j+1}} f(\tau) d\tau \Bigg) \int\limits_0^t \beta_j
\end{align}
and check the the relation $dh + hd = id - P_Z$ for 1-forms:
\begin{align}
    \begin{split}
        (dh + hd)(f(t)dt) & = d \Bigg[ \int\limits_0^t f(\tau) d\tau - \sum\limits_j \Bigg( \int\limits_{t_j}^{t_{j+1}} f(\tau) d\tau \Bigg) \int\limits_0^t \beta_j \Bigg] = \\
        & = f(t)dt - \sum\limits_j \Bigg( \int\limits_{t_j}^{t_{j+1}} f(\tau) d\tau \Bigg) \beta_j = (id - P_Z) f(t)dt \\
    \end{split}
\end{align}
and for 0-forms:
\begin{align}
    \begin{split}
        (dh + hd)(g(t)) & = \int\limits_0^t g'(\tau) d\tau - \sum\limits_j \Bigg( \int\limits_{t_j}^{t_{j+1}} g'(\tau) d\tau \Bigg) \int\limits_0^t \beta_j = \\
        & = g(t) - g(0) - \sum\limits_j (g(t_{j+1}) - g(t_j)) \int\limits_0^t \beta_j = \\
        & = g(t) - g(0) - g(0) \Bigg( \int\limits_0^t(\beta_{n-1} - \beta_0) \Bigg) - \\
        & - \sum\limits_{j \ne 0} g(t_j) \Bigg( \int\limits_0^t(\beta_{j-1} - \beta_j) \Bigg) = \\
        & = g(t) - \sum\limits_j g(t_j) \alpha_j = (id - P_Z)g(t)
    \end{split}
\end{align}

The induced action (\ref{induced_action}) is analogous to the case of the interval
\begin{equation}
    S^{ind} = \langle \varphi^*_Z, d\varphi_Z \rangle + (-1)^{|\varphi_Z^*|} \langle \varphi^*_Z, \mathcal{L}_{v_c} \varphi_Z \rangle + \langle \varphi^*_Z, \mathcal{L}_{v_c} h \mathcal{L}_{v_c} \varphi_Z \rangle + \frac{1}{2} f_{ab}^d c^a c^b c_d^*
\end{equation}

We introduce notations
\begin{align}
    & \nu_c(t_j) = \nu_c^j \\
    & \nu'_c(t_j) = \nu'^j_c
\end{align}

Finally, let us compute the action
\begin{align}
    & d\alpha_j = \beta_{j-1} - \beta_j \\
    & \langle \beta^i, d\alpha_j \rangle = \delta_{j-1}^i - \delta_j^i \\
    & \langle \alpha^i, \mathcal{L}_{v_c} \alpha_j \rangle = \nu_c^i (\beta_{j-1}^i - \beta_j^i) \\
    & \langle \beta^i, \mathcal{L}_{v_c} \beta_j \rangle = \int\limits_{t_i}^{t_{i+1}} d \Big( \nu_c(t) \vartheta_j(t) \Big) = \nu_c^{i+1} \beta_j^{i+1} - \nu_c^i \beta_j^i \\
    \begin{split}
        h \mathcal{L}_{v_c} \beta_j & = \int\limits_0^t d \Big( \nu_c \vartheta_j \Big) - \sum\limits_m \Bigg( \int\limits_{t_m}^{t_{m+1}} d \Big( \nu_c \vartheta_j \Big) \Bigg) \int\limits_0^t \beta_m = \\
        & = \nu_c(t) \vartheta_j(t) - \nu_c^0 \beta_j^0 - \sum\limits_m \Big( \nu_c^{m+1} \beta_j^{m+1} - \nu_c^m \beta_j^m \Big) \int\limits_0^t \beta_m = \\
        & = \nu_c(t) \vartheta_j(t) - \sum\limits_m \nu_c^m \beta_j^m \alpha_m(t) \\
    \end{split} \\
    \begin{split}
        \mathcal{L}_{v_c} h \mathcal{L}_{v_c} \beta_j & = \nu_c(t) \frac{\partial}{\partial t} \Big( \nu_c(t) \vartheta_j(t) - \sum\limits_m \nu_c^m \beta_j^m \alpha_m(t) \Big) = \\
        & = \nu_c(t) \nu'_c(t) \vartheta_j(t) + \nu_c(t) \nu_c(t) \vartheta'_j(t) - \\
        & - \nu_c(t) \sum\limits_m \nu_c^m \beta_j^m (\vartheta_{m-1}(t) - \vartheta_m(t)) \\
    \end{split} \\
    & \langle \alpha^i, \mathcal{L}_{v_c} h \mathcal{L}_{v_c} \beta_j \rangle = \nu_c^i \nu'^i_c \beta_j^i - \nu_c^i \sum\limits_m \nu_c^m \beta_j^m (\beta_{m-1}^i - \beta_m^i)
\end{align}

Thus, we get the $L_\infty$ module structure
\begin{align}
    \begin{split}
        S^{ind} & = \psi_i^* (\delta_{j-1}^i - \delta_j^i) \varphi^j - \varphi_i^* \nu_c^i (\beta_{j-1}^i - \beta_j^i) \varphi^j + \psi_i^* (\nu_c^{i+1} \beta_j^{i+1} - \nu_c^i \beta_j^i) \psi^j + \\
        & + \varphi_i^* \Big( \nu_c^i \nu'^i_c \beta_j^i - \nu_c^i \sum\limits_m \nu_c^m \beta_j^m (\beta_{m-1}^i - \beta_m^i) \Big) \psi^j + \frac{1}{2} f_{ab}^d c^a c^b c_d^*
    \end{split}
\end{align}

\section{Square}

In this section we consider the manifold $X = [0, 1] \times [0, 1]$ with the coordinates $x, y$. The vector field $v_c$ takes the form
\begin{align}
    & v_c(x,y) = \nu_c^x(x,y) \frac{\partial}{\partial x} + \nu_c^y(x,y) \frac{\partial}{\partial y}, \\
    & \nu_c^x = \sum\limits_{n,m} c^{x,nm} x^n y^m, \\
    & \nu_c^y = \sum\limits_{n,m} c^{y,nm} x^n y^m
\end{align}

The triangulation is given by
\begin{gather}
    \gamma^* = [0, 1] \times [0, 1] \\
    \beta^0 = [(0,0), (1,0)], \beta^1 = [(1,0), (1,1)], \beta^2 = [(1,1), (0,1)], \beta^3 = [(0,1), (0,0)] \\
    \alpha^0 = (0,0), \alpha^1 = (1,0), \alpha^2 = (1,1), \alpha^3 = (0,1) \\
    Z = Span(\alpha^0,..., \alpha^3, \beta^0,..., \beta^3, \gamma^*)
\end{gather}

Let us define the constructions that willl be used in further calculations: 2-form $\gamma = dx dy$; 1-forms:
\begin{align}
    & \beta_0 = \Bigg( \int\limits_y^1 d\Tilde{y} \Bigg) dx = (1 - y) dx \\
    & \beta_1 = \Bigg( \int\limits_0^x d\Tilde{x} \Bigg) dy = x dy \\
    & \beta_2 = -\Bigg( \int\limits_0^y d\Tilde{y} \Bigg) dx = -y dx \\
    & \beta_3 = -\Bigg( \int\limits_x^1 d\Tilde{x} \Bigg) dy = -(1 - x) dy;
\end{align}
and functions:
\begin{align}
    & \alpha_0 = 1 + \int (\beta_3 - \beta_0) = (1 - x)(1 - y)\\
    & \alpha_1 = \int (\beta_0 - \beta_1) = x(1 - y) \\
    & \alpha_2 = \int (\beta_1 - \beta_2) = xy \\
    & \alpha_3 = \int (\beta_2 - \beta_3) = y(1 - x).
\end{align}

Since $d\beta_j = \gamma$, forms $\beta_i - \beta_j$ are closed, and the functions $\alpha_i$ are defined correctly. Note that $\langle \gamma^*, \gamma \rangle = 1, \langle \beta^i, \beta_j \rangle = \delta_j^i, \langle \alpha^i, \alpha_j \rangle = \delta_j^i$. Then the space dual to the space of chains is
\begin{align}
    \Omega_Z = Span(\alpha_0,..., \alpha^3, \beta_0,..., \beta^3, \gamma)
\end{align}

The fields under consideration are
\begin{align}
    & \varphi_Z = \sum\limits_{i=0}^3 \alpha_i \varphi^i + \sum\limits_{j=0}^3 \beta_j \psi^j + \gamma \omega \\
    & \varphi_Z^* = \sum\limits_{i=0}^3 \varphi_i^* \alpha^i + \sum\limits_{j=0}^3 \psi_j^* \beta^j + \omega^* \gamma^*
\end{align}

We introduce the projector $P_Z$ ($g$, $f$ and $e$ are 0, 1 and 2 -forms respectively) analogously to the previous sections
\begin{align}
    & P_Z(g) = \sum\limits_j \langle \alpha^j, g \rangle \alpha_j \\
    & P_Z(f) = \sum\limits_j \langle \beta^j, f \rangle \beta_j \\
    & P_Z(e) = \langle \gamma^*, e \rangle \gamma
\end{align}
To define the homotopy consider the operator $I:\Omega_X \to \Omega_X$; for 2-forms:
\begin{align}
    I(e(x,y) dx dy) = \frac{1}{2} \Bigg( \int\limits_0^x e(\Tilde{x}, y) d\Tilde{x} \Bigg) dy - \frac{1}{2} \Bigg( \int\limits_0^y e(x, \Tilde{y}) d\Tilde{y} \Bigg) dx
\end{align}
for 1-forms:
\begin{align}
\begin{split}
    I(f_x(x,y) dx + f_y(x,y) dy) = & \frac{1}{2} \Bigg( \int\limits_0^x f_x(\Tilde{x}, 0) d\Tilde{x} + \int\limits_0^y f_y(x, \Tilde{y}) d\Tilde{y} + \\
    & + \int\limits_0^y f_y(0, \Tilde{y}) d\Tilde{y} + \int\limits_0^x f_x(\Tilde{x}, y) d\Tilde{x} \Bigg)
\end{split}
\end{align}
The last expression is simply the sum of the integrals of an 1-form $f$ along the edges of the rectangle with vertices $(0,0), (0,x), (0,y), (x,y)$. For 0-forms $g(t)$ $I(g) = 0$. Using $I$, we can now define the homotopy (e and f are 2- and 1- forms respectively):
\begin{align}
    & h(e) = I(e) - \sum\limits_i \langle \beta^i, I(e) \rangle \beta_i \\
    & h(f) = I \Big( f - \sum\limits_i \langle \beta^i, f \rangle \beta_i \Big)
\end{align}

Let us check the condition (\ref{homotopy_cond}) in the form $dh + hd = id - P_Z$:\begin{align}
    \begin{split}
        (dh + hd) g(x,y) & = h(dg) = I \Big( dg - \sum\limits_i \langle \beta^i, dg \rangle \beta_i \Big) = \\
        & = g(x,y) - g(0,0) - I \Big( \sum\limits_i \big( \langle \alpha^{i+1}, g \rangle - \langle \alpha^i, g \rangle \big) \beta_i \Big) = \\
        & = g(x,y) - g(0,0) - \sum\limits_i \langle \alpha^i, g \rangle I(\beta_{i-1} - \beta_i ) = \\
        & = g(x,y) - \sum\limits_i \langle \alpha^i, g \rangle \alpha_i = (id - P_Z) g \\
    \end{split} \\
    \begin{split}
        (dh + hd) e & = d(h(e)) = d \Big( I(e) - \sum\limits_i \langle \beta^i, I(e) \rangle \beta_i \Big) = \\
        & = e - \Big( \sum\limits_i \langle \beta^i, I(e) \rangle \Big) \gamma = e - \langle \gamma^*, e \rangle \gamma = (id - P_Z) e \\
    \end{split} \\
    & h d f = h d (f_x dx + f_y dy) = h(df) = I(df) - \sum\limits_i \langle \beta^i, I(df) \rangle \beta_i \\
    \begin{split}
        2 I(df) & = \Bigg( f_x(x, y) - f_x(x, 0) - \int\limits_0^y \frac{\partial f_y}{\partial x} d\tilde{y} \Bigg) dx + \\
        & + \Bigg( f_y(x, y) - f_y(0, y) - \int\limits_0^x \frac{\partial f_x}{\partial y} d\tilde{x} \Bigg) dy \\
    \end{split} \\
    & \sum\limits_i \langle \beta^i, I(df) \rangle \beta_i = \Big( \sum\limits_i \langle \beta^i, f \rangle \Big) \frac{\beta_1 + \beta_2}{2} = \langle \gamma^*, df \rangle I(\gamma) \\
    \begin{split}
        2 d I(f) & = f_x(x, y) dx + f_y(x, y) dy + f_x(x, 0) dx + f_y(0, y) dy + \\
        & + \Bigg( \int\limits_0^y \frac{\partial f_y}{\partial x} d\tilde{y} \Bigg) dx + \Bigg( \int\limits_0^x \frac{\partial f_x}{\partial y} d\tilde{x} \Bigg) dy \\
    \end{split} \\
    & I(df) + d I(f) = f \\
    \begin{split}
        (hd + dh) f & = I(df) + d I(f) - \sum\limits_i \langle \beta^i, I(df) \rangle \beta_i - \sum\limits_i \langle \beta^i, f \rangle d I(\beta_i) = \\
        & = f - \sum\limits_i \langle \beta^i, f \rangle \beta_i - \langle \gamma^*, df \rangle I(\gamma) + \sum\limits_i \langle \beta^i, f \rangle I(d\beta_i) = \\
        & = (1 - P_Z) f + \sum\limits_i \langle \beta^i, f \rangle I(\gamma) - \langle \gamma^*, df \rangle I(\gamma) = (1 - P_Z) f \\
    \end{split}
\end{align}
    
Let us compute the components of action
\begin{align}
    & d\alpha_j = \beta_{j-1} - \beta_j \label{square_begin} \\
    & \langle \beta^i, d\alpha_j \rangle = \delta_{j-1}^i - \delta_j^i \\
    & d\beta_j = \gamma \\
    & \langle \gamma^*, d\beta_j \rangle = 1 \\
    \begin{split}
        \langle \alpha^i, \mathcal{L}_{v_c} \alpha_j \rangle & = \langle \alpha^i, \nu_c^x \frac{\partial}{\partial x} \int (\beta_{j-1} - \beta_j) + \nu_c^y \frac{\partial}{\partial y} \int (\beta_{j-1} - \beta_j) \rangle \\
        & = \nu_i (\beta_{j-1}^i - \beta_j^i) \\
    \end{split} \\
    \begin{split}
        \langle \beta^i, \mathcal{L}_{v_c} \beta_j \rangle & = \langle \beta^i, d (v_c \beta_j) + i_{v_c} d\beta_j \rangle = \nu_{i+1} \beta_j^{i+1} - \nu_i \beta_j^i + \\
        & + \int\limits_{\beta^i} \nu_c^x dy - \nu_c^y dx \\
    \end{split} \\
    & \langle \gamma^*, \mathcal{L}_{v_c} \gamma \rangle = \langle \gamma^*, d (\nu_c^x dy - \nu_c^y dx) \rangle = \int\limits_{[0, 1]^2} \mathbf{div}(v_c) dx dy \\
    \begin{split}
        \langle \alpha^i, \mathcal{L}_{v_c} h \mathcal{L}_{v_c} \beta_j \rangle & = \langle \alpha^i, \mathcal{L}_{v_c} h \big( d (v_c \beta_j) + i_{v_c} d\beta_j \big) \rangle = \langle \alpha^i, \mathcal{L}_{v_c} (hd + dh) (v_c \beta_j) \rangle + \\
        & + \langle \alpha^i, \mathcal{L}_{v_c} h \big( \nu_c^x dy - \nu_c^y dx \big) \rangle = \langle \alpha^i, \mathcal{L}_{v_c} (v_c \beta_j - \sum\limits_m \nu_m \beta_j^m \alpha_m) \rangle + \\
        & + \langle \alpha^i, \mathcal{L}_{v_c} h \big( \nu_c^x dy - \nu_c^y dx \big) \rangle \\
        & = \nu_{x,i} \frac{\partial (\nu_i \beta_j)}{\partial x} + \nu_{y,i} \frac{\partial (\nu_i \beta_j)}{\partial y} - \sum\limits_m \nu_i (\beta_{m-1}^i - \beta_m^i) \nu_m \beta_j^m + \\
        & + \frac{1}{2} \langle \alpha^i, \nu_c^x(x, y) \nu_c^x(x, 0) - \nu_c^x \int\limits_0^y \frac{\partial \nu_c^y}{\partial x}(x, \tilde{y}) d\title{y} \rangle + \\
        & + \frac{1}{2} \langle \alpha^i, -\nu_c^y(x, y) \nu_c^y(0, y) + \nu_c^y \int\limits_0^y \frac{\partial \nu_c^x}{\partial x}(\tilde{x}, y) d\title{x} \rangle - \\
        & - \langle \alpha^i, \sum\limits_k \Big( \nu_c^x \frac{\partial I(\beta_k)}{\partial x} + \nu_c^y \frac{\partial I(\beta_k)}{\partial y} \Big) \int\limits_{\beta^k} \nu_c^x dy - \nu_c^y dx \rangle \\
    \end{split}
\end{align}
Denote
\begin{align}
    \begin{split}
        \xi := \mathcal{L}_{v_c} h \mathcal{L}_{v_c} \gamma & = \mathcal{L}_{v_c} h d(\nu_c^x dy - \nu_c^y dx) = \mathcal{L}_{v_c} h \big( \mathbf{div}(v_c) dx dy \big) \\
        & = \frac{1}{2} \Bigg( d \Big( \nu_c^y \big(\nu_c^x(x, y) - \nu_c^x(0, y) + \int\limits_0^x \frac{\partial \nu_c^y}{\partial y} d\tilde{x} \big) - \\
        & - \nu_c^x \big( \nu_c^y(x, y) - \nu_c^y(x, 0) + \int\limits_0^y \frac{\partial \nu_c^x}{\partial x} d\title{y} \big) \Big) + \mathbf{div}(v_c) \Big( \nu_c^x dy - \nu_c^y dx \Big) - \\
        & - \Big( \int\limits_{[0, 1]^2} \mathbf{div}(v_c) dx dy \Big) \Big( d \big( v_c (\beta_1 + \beta_2) \big) + 2 (\nu_c^x dy - \nu_c^y dx) \Big) \Bigg) \\
    \end{split}
\end{align}
The most complicated term (cubic in $c^a$) has the form
\begin{align}
    \begin{split}
        \langle \alpha^i, \mathcal{L}_{v_c} h \mathcal{L}_{v_c} h \mathcal{L}_{v_c} \gamma \rangle & = \langle \alpha^i, \mathcal{L}_{v_c} h \xi \rangle = \langle \alpha^i, \mathcal{L}_{v_c} \Big( I(\xi) - \sum\limits_i I(\beta_i) \int\limits_{\beta^i} \xi \Big) \rangle \label{square_end} \\
    \end{split}
\end{align}

Thus, from (\ref{square_begin})-(\ref{square_end}) we get the $L_\infty$ module structure
\begin{align}
    \begin{split}
        S^{ind} & = \psi_i^* \langle \beta^i, d\alpha_j \rangle \varphi^j + \omega^* \langle \gamma^*, d\beta_j \rangle \psi^j - \varphi_i^* \langle \alpha^i, \mathcal{L}_{v_c} \alpha_j \rangle \varphi^j + \psi_i^* \langle \beta^i, \mathcal{L}_{v_c} \beta_j \rangle \psi^j - \\
        & - \omega^* \langle \gamma^*, \mathcal{L}_{v_c} \gamma \rangle \omega + \varphi_i^* \langle \alpha^i, \mathcal{L}_{v_c} h \mathcal{L}_{v_c} \beta_j \rangle \psi^j + \psi_i^* \langle \beta^i, \mathcal{L}_{v_c} h \mathcal{L}_{v_c} \gamma \rangle \omega + \\
        & + \varphi_i^* \langle \alpha^i, \mathcal{L}_{v_c} h \mathcal{L}_{v_c} h \mathcal{L}_{v_c} \gamma \rangle \omega + \frac{1}{2} f_{ab}^d c^a c^b c_d^*
    \end{split}
\end{align}

\section*{References}

\printbibliography[heading=none]

\end{document}